\documentclass{moriond}

\usepackage[table]{xcolor}
\usepackage{amsmath}
\usepackage{amssymb}
\usepackage{wrapfig}

\def\be{\begin{equation}}
\def\ee{\end{equation}}
\def\bea{\begin{eqnarray}}
\def\eea{\end{eqnarray}}

\def \MET{\rm E{\!\!\!/}_T}

\newcommand{\opdhchi}{\mathcal{O}_{DH\chi\chi}}

\usepackage[normalem]{ulem}
\usepackage{xcolor}

\newcommand{\Photo}{}

\begin{document}
	\vspace*{4cm}
	\title{A Window onto New Invisible Particles via Semi-Visible Higgs Decays}
	
	\def\BNL{High Energy Theory Group, Physics Department, Brookhaven National Laboratory, Upton, NY 11973, USA}
	\def\MON{
		School of Physics and Astronomy, Monash University, Wellington Road, Clayton, Victoria 3800, Australia}
	
	\author{Sally Dawson}
	\address{\BNL}
	\author{Arnab Roy and German Valencia}
	\address{\MON}

	\maketitle
	\abstracts{
		Searches for new physics continue at the LHC in several forms, including new-particle searches, precision measurements of SM couplings, and searches for signals of new invisible particles. In this talk, we discuss the reach in parameter space of new invisible particles with the semi-visible Higgs decay modes $H\to \ell^+\ell^- + ~\rm E{\!\!\!/}_T$ and $H\to jj + ~\rm E{\!\!\!/}_T$. We first parametrise the new invisible particles and their interactions with the SM through an effective field theory at dimension six. We then study the respective signals and the corresponding background, finding small signals with large backgrounds. We find, however, that the kinematics of these processes are sufficiently rich to allow a signal extraction that we first quantify with a cut-based analysis and later with a multivariate BDT.}
	
	Decays into invisible particles that show up as missing energy have a rich tradition, with a notable example being the counting of neutrino species with the measurement of the invisible $Z$ width. Higgs decays with missing energy are sensitive, for example, to new weakly interacting particles, to additional neutral states beyond neutrinos, to hidden sectors that couple weakly to known particles, and to dark matter. In Table~\ref{tab:exp}, we collect some possible Higgs decay modes with $\MET$, with the ones highlighted in red being the subject of this talk, the semi-visible modes. The SM rates are taken from \cite{LHCHiggsCrossSectionWorkingGroup:2016ypw,dEnterria:2023wjq}, the current limits from \cite{ParticleDataGroup:2024cfk}, and the HL-LHC projections from \cite{Cepeda:2019klc}.
	\begin{table}[hbt] 
		\caption{Some Higgs decay modes with $\MET$ indicating the SM contribution and the current experimental limits and HL-LHC projections when available. Charged leptons include $\ell=e,\mu$.}
		\centering
		\begin{tabular}{|c|c||c|c|c|}\hline
			Mode & SM process & SM BR& current limit ($\%$)&HL-LHC ($\%$)\\ \hline
			$H\to$~invisible & $H\to ZZ^*\to\nu\bar\nu \nu\bar\nu$ &  $1.06\times 10^{-3} $ & $<10.7$ & 2-3\\
			$H\to\gamma$~invisible & $H\to Z\gamma\to\nu\bar\nu~ \gamma $&  $3.6\times 10^{-4}$& $<1.3$ &-\\ \hline 
			\cellcolor{red!30} $H\to\ell^+\ell^-$~invisible & $H\to \ell^+\ell^- \nu\bar\nu$&  $1.06\times 10^{-2}$ & -& $10^{-3}$ \\
			\cellcolor{red!30} $H\to jj$~invisible& $H\to qq \nu\bar\nu$&  $7.5\times 10^{-3}$& -&-\\
			\hline
		\end{tabular}
		\label{tab:exp}
	\end{table}
	
	Previous work in \cite{Aguilar-Saavedra:2022xrb} considered models, including ALPs, leading to $h\to\ell^+\ell^-X$, where  $X$ is a single invisible particle. They found sensitivity to unexplored regions of parameter space up to an irreducible neutrino floor corresponding to the SM process $H\to ZZ^*\to\ell^+\ell^- +\nu{\overline{\nu}}$ (dubbed $ZH$-lep below in our listing of background). There are many models where the $\MET$ is due to more than one missing particle, e.g. \cite{Guchait:2020wqn,Dey:2023exa}, and here we consider those where the invisible particle is the only new one below the EW scale, and can be described with DSMEFT \cite{Criado:2021trs,Aebischer:2022wnl,He:2022ljo}.
	
	From these lists, we select the dimension six operators that appear in our study with new invisible states appearing as complex scalars or Dirac fermions (which also cover the cases of real scalars with a $\mathbb{Z}_2$ symmetry and of Majorana fermions). This leads to the following
	\begin{align}
	{\mathcal O}_{DH\partial\phi} &= (H^\dagger \overleftrightarrow{D}^\mu H)(\phi^\dagger \overleftrightarrow{\partial_\mu}\phi)  
	& {\mathcal O}_{e\phi}{}  &= (\overline l_i e_j H) \phi^\dagger\phi \\
	{\mathcal O}_{u\phi}{} &= (\overline q_i u_j \widetilde H) \phi^\dagger\phi &  {\mathcal O}_{d\phi}{} &= (\overline q_i\, d_j H) \phi^\dagger\phi\\
	{\mathcal O}_{DH\chi\chi}{} &=  (iH^\dagger \overleftrightarrow{D}^\mu H)  (\overline \chi\gamma_\mu  \chi) &
	{\mathcal O}_{DH\chi\chi 2}{} &=  (iH^\dagger \overleftrightarrow{D}^\mu H)  (\overline \chi\gamma_\mu\gamma^5  \chi).
	\label{eq:lag}
	\end{align}
	As our only observable is a CP-even rate, and we treat all Standard Model (SM) fermions appearing in the final state as massless, there is no need to consider CP-violating or flavour-changing operators separately. The list of operators in Eq.~\ref{eq:lag} omits those involving only the Higgs and new invisible fields. Their contributions necessarily involve light fermion masses, and once constraints from the Higgs invisible width \cite{Criado:2021trs,Dawson:2025dmi} are taken into account, they are unobservable in semi-visible Higgs decays.
	Finally, the kinematic window for semi-visible decays is $m_{\chi,\phi}\lesssim m_h/2$, about 50~GeV in practice.
	
\begin{figure}
		\centering
		\includegraphics[width=0.7\linewidth]{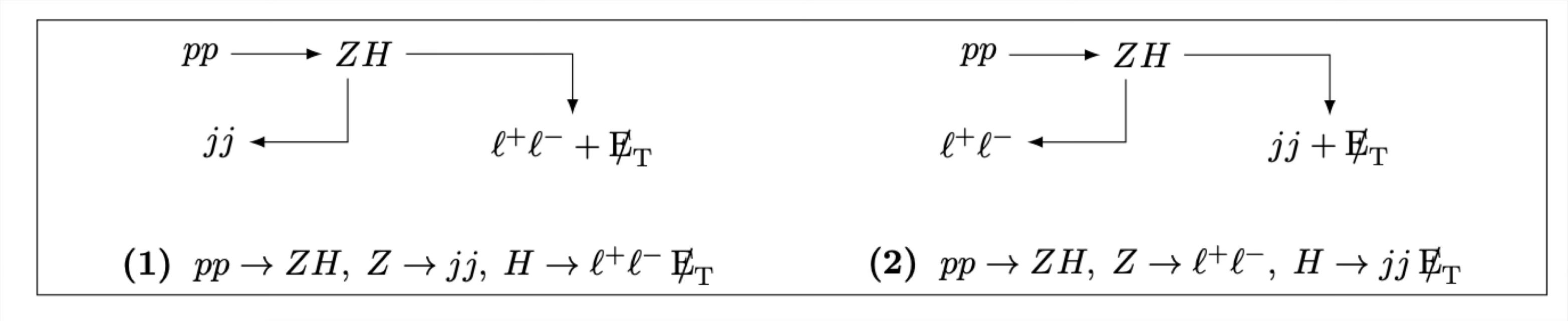}
		\caption{Semivisible Higgs decay: leptonic mode with tagging $Z\to jj$ (left), and hadronic mode with tagging $Z\to \ell^+\ell^-$ (right).}
		\label{fig:processes}
	\end{figure}

\begin{figure}[h]
		\centering
		\includegraphics[width=1\linewidth]{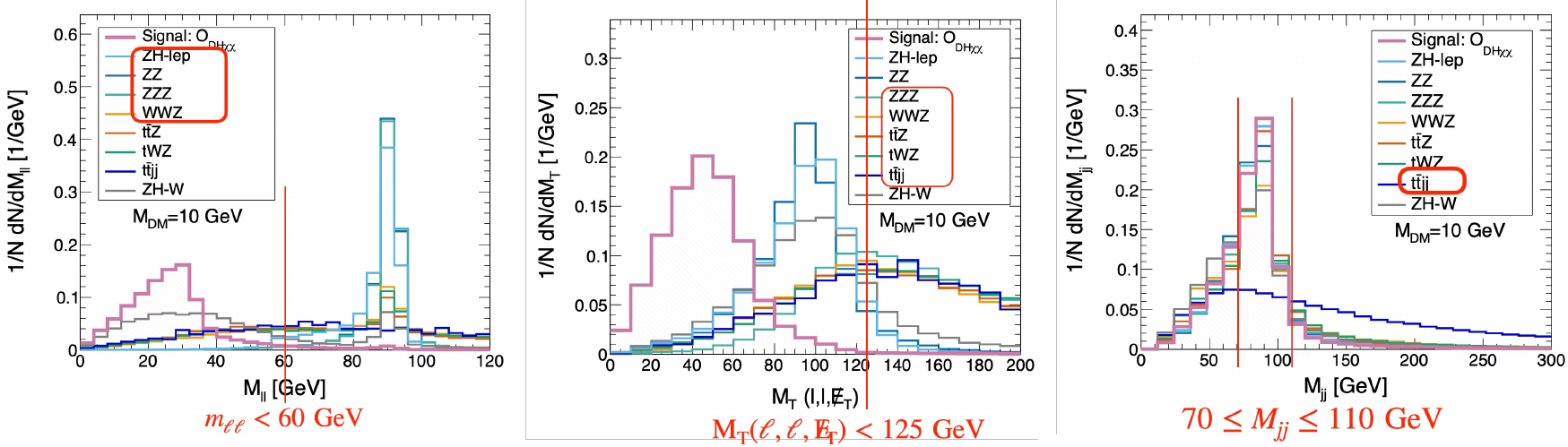}
		\caption{Comparison of selected kinematic distributions for a signal from the operator $\opdhchi$ and different background processes. In each panel, we have circled the background processes that can be reduced with the given cut. For the signal, we use $C_{DH\chi\chi}=1,~\Lambda=1$ TeV, and show separately normalised processes for $\sqrt{s}=14~{\textrm{TeV}}$.}
		\label{fig:cuts}
	\end{figure}

	The Higgs is produced at the LHC through gluon fusion, vector boson fusion, and as $ZH$, which successively have lower cross-section but also less background. For this first study, we chose the $ZH$ production mode as it is very clean, and the additional $Z$ helps tag the decay. We consider the two processes sketched in Figure~\ref{fig:processes}. Notice that both processes have the same final state, but the cuts that separate them from background requiring the invariant mass of the fermion pair from the tagging $Z$ to be near the $Z$ mass and the invariant mass of the other fermion pair to be {\it well below} the $Z$ mass,  also separate them from each other, the resulting spillover between them being of order 1\%.
	The signal is implemented via the usual workflow of \texttt{FeynRules}~\cite{Alloul:2013bka}, \nolinkurl{Madgraph5}~\cite{Alwall:2014hca} at LO, with the \texttt{PDF4LHC15\_nlo\_mc}, \nolinkurl{Pythia8}~\cite{Sjostrand:2006za}, and \nolinkurl{Delphes3}~\cite{deFavereau:2013fsa}, configured with the CMS detector card. At the \nolinkurl{Madgraph5} level we introduce the following background processes: $ZH$-lep:  $H\to ZZ^*\to\ell^+\ell^- +\nu{\overline{\nu}}, Z\to jj$, $ZZ$: $Z\to jj,~Z\to \ell^+\ell^-$, $ZZZ$: $Z\to jj,~Z\to \ell^+\ell^-,~Z\to \nu{\overline{\nu}}$, $WWZ$: $ \to jj\,\ell^+\,\ell^+\,\nu\,{\overline{\nu}}$, $t\bar t Z$: $ \to jj\,\ell^+\,\ell^-\,\nu\,{\overline{\nu}} \,b\,\bar{b}$, $tWZ$: $ \to jj\,\ell^+\,\ell^-\,\nu\,{\overline{\nu}}$, $ZH-W$: $H\to WW^*\to\ell^+\ell^- +\nu{\overline{\nu}}, Z\to jj$, and $t\bar{t}-inclusive:~ ~t\to b\,\ell^+\nu,\; \bar{t}\to \bar{b}\,\ell^-\bar{\nu}$. We first carry out a standard cut analysis for the semi-visible leptonic case to gain some insight into the effect of the different cuts. Three of the most important ones are shown in Figure~\ref{fig:cuts}, which highlights which ones help to reduce which background.
	
	As already mentioned, this is a process with a small signal and large background, as summarised in Table~\ref{tab:sigmas}.\footnote{This table has been updated from the original presentation to correct erroneous entries.}
	\begin{table}[hbt] 
		\caption{Top panel: background cross-sections for LHC at $\sqrt{S}=14$ TeV with basic selection cuts for 
			two leptons with $p_T(\ell_1)>20~{\rm GeV}, ~p_T(\ell_2)>10~~{\rm GeV}, |\eta|_\ell < 2.5$ and two jets with $p_T(j)>20~{\rm GeV}, ~ |\eta|_j <4.0$, and the cumulative effect of all analysis cuts $\rm M_{\ell\ell}<60~{\rm GeV},~\rm M_T(\ell,\ell,\rm E{\!\!\!/}_T)<125~ {\rm GeV},~\rm E{\!\!\!/}_T>20.0\; {\rm GeV},~70< M_{jj} <110~\rm{GeV}~+~{\rm N}(\rlap{/}b-jet)=2$ and $b$-jet vetoing. Bottom panel signal cross-sections for different operators and $M_{\phi,\chi}=10~{\rm GeV}$.}
		\centering
		\resizebox{\textwidth}{!}{%
			\begin{tabular}{|c|c|c|c|c|c|c|c|c|}
				\hline   & ZH-lep  & ~~ZH-W~~  & ~~~ZZ~~~  & ZZZ  & ~~WWZ~~  & ~~$\rm t\bar{t}Z$~~  & ~~tWZ~~ & ~~$\rm~t\bar{t}-inc$~~  \\
				\hline Cross-section (fb)  & $0.05$  & $0.06$  & $547$  & $0.33$  & $2.2$  & $17.7$  & $6.7$  & $45719$\\ 
				\hline
				\hline after all cuts  & {$2.2 \times 10^{-4}$}  & {$0.002$}  & {$0.09$}  & {$2 \times 10^{-4}$}  & {$0.016$}  & {$0.0065$}  & {$0.006$}  & $1.4$\\ 
				\hline
		\end{tabular}}    
		\centering
		\resizebox{0.7\textwidth}{!}{
			\begin{tabular}{|c|c|c|c|c|}
				\hline
				& $C_{DH\chi\chi}=10$
				& $C_{DH\chi\chi2}=10$ & $C_{DH\partial\phi}=10$ & $C_{e\phi}=10$ \\
				\hline
				Cross-section (fb)  & $0.021$ & $0.019$  & $0.005$ & $0.004$ \\
				\hline
				after all cuts
				& {$0.004$}  & {$0.004$} & {$0.001$} & {$0.0003$} \\
				\hline
		\end{tabular}   }   
		\label{tab:sigmas}
	\end{table}
	With 3000~fb$^{-1}$, after all cuts we find  $\sigma_{bkg}=1.52~\rm fb$, implying that we need about  207 signal events for a $3\sigma$ significance, which results in a reach of 
	\begin{align}
	C_{DH\chi\chi} \lesssim  41{\rm~TeV}^{-2},~~
	C_{DH\partial\phi} \lesssim  83{\rm~TeV}^{-2},~~
	C_{e\phi} \lesssim  152{\rm~TeV}^{-2}
	\end{align}
	
\begin{figure}
		\centering
		\includegraphics[width=1\linewidth]{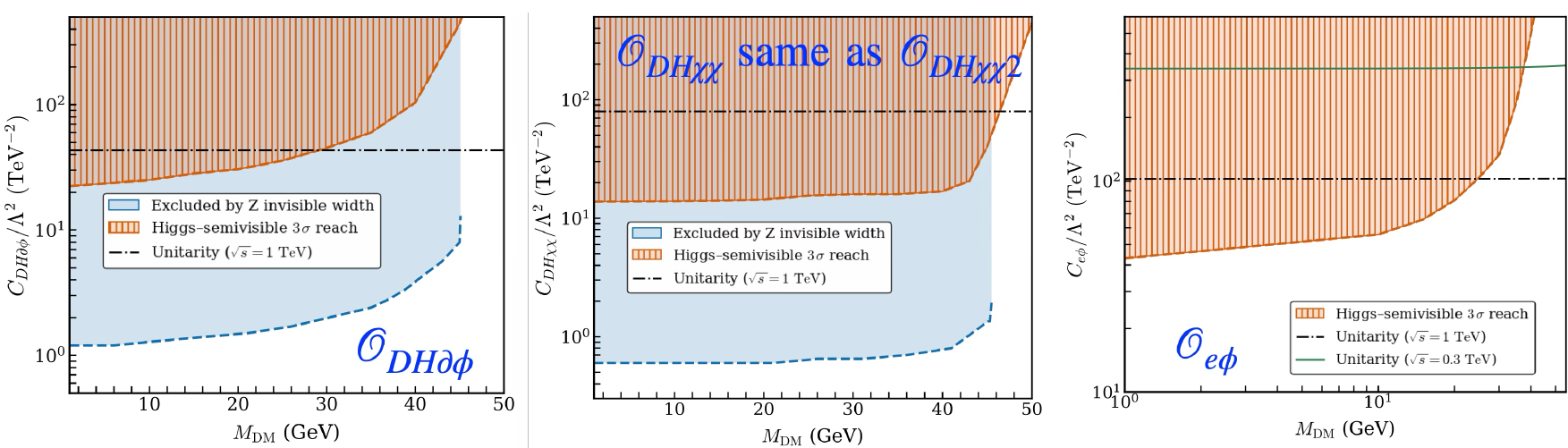}
		\caption{Parameter space that can be excluded by semi-visible Higgs decay shown in brown. For derivative operators, the invisible $Z$ width (shown in blue) provides a more stringent constraint.}
		\label{fig:parameters}
	\end{figure}

	To improve this analysis with a multivariate BDT, we start from a sample of events {\it after} $b$-jet vetoing and requiring two leptons, reducing the original $t\bar t-inc$ background down to 1.35\%. Multivariate classification is then performed on this sample using a boosted decision tree implemented in TMVA. (Toolkit for Multivariate Analysis). Some additional variables are included in the classifier, $\Delta \phi (\ell,\ell), \Delta \phi (\ell_1,\rm E{\!\!\!/}_T), \Delta \phi (\ell_2,\rm E{\!\!\!/}_T)$. We then use 
	60\% of the generated events to train the BDT and 40\% for testing, repeating for each operator and for a few DM mass values. 
	The BDT response distributions for the training and testing samples are in good agreement. The Kolmogorov–Smirnov test is performed to verify that the classifier is not overtrained. The resulting variable ranking and the signal significance as a function of the classifier output can be found in \cite{Dawson:2025dmi}. The final results, obtained with a cut in the classifier at 0.997, reflect a significant improvement in significance compared to the cut-based analysis, and are shown in Figure~\ref{fig:parameters}.
	The figure shows the parameter space that can be excluded by semi-visible Higgs decay in brown. It indicates that for derivative operators, the invisible $Z$ width (shown in blue) provides a more stringent constraint. As the absolute numbers of the allowed WC appear large, we also compare them to their unitarity limits. A limit shown assumes a new physics scale $\sim 3$~TeV and requires unitarity to hold up to 1~TeV. For the Yukawa-like operators, we include a second unitarity limit with a new physics scale near 1~TeV, where unitarity is only required to hold up to about 300~GeV. Note that this is sufficient to describe Higgs decay processes, where the energies probed are below the Higgs mass.
	
	\begin{wrapfigure}{l}{0.45\textwidth}
		\centering
		\includegraphics[width=1\linewidth]{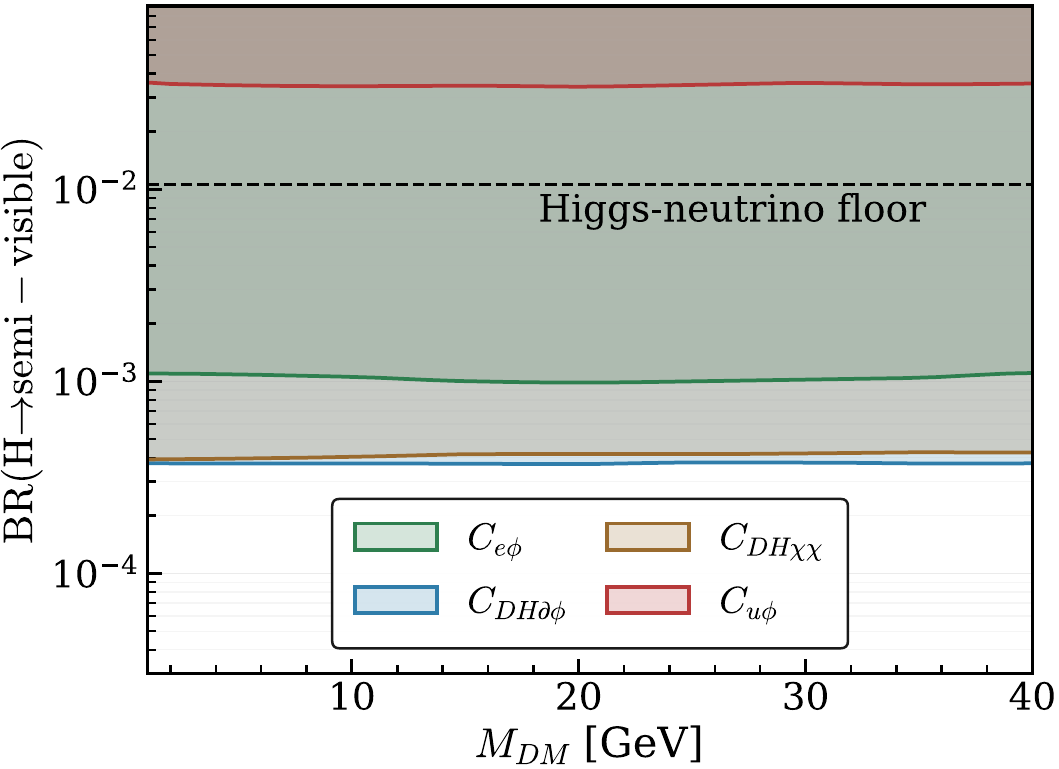}
		\caption{$3\sigma$ sensitivities for HL-LHC in terms of branching fractions. }
		\label{fig:bratios}
	\end{wrapfigure}
	In Figure~\ref{fig:bratios}, we show the $3\sigma$ sensitivities of HL-LHC, 14~TeV, 3000~fb$^{-1}$ to the different Wilson coefficients, in units of the implied Higgs branching fraction. The figure also illustrates how the “Higgs neutrino floor” in this case is reducible. The dotted black line corresponds to $BR(H \to \ell^+\ell^-\nu\bar{\nu})_{SM} = 1.06\times 10^{-2}$, or the SM contribution to semi-visible Higgs leptonic decay. The corresponding line for the hadronic case (relevant for the quark operators) is about 30\% lower, not shown. Not shown are two couplings with almost identical limits, $C_{u\phi}\sim C_{d\phi}$ and $C_{DH\chi\chi}\sim C_{DH\chi\chi 2}$.
	
	In summary, Higgs semi-visible searches complement the Higgs invisible searches for the kinematic window $m_{\phi,\chi}\lesssim 50$ GeV. This is a
	process with a small signal and large backgrounds, but with kinematics sufficiently rich that a meaningful signal can be extracted at HL-LHC, in particular, making the “Higgs neutrino floor” reducible. Monojet searches provide a much stronger constraint for Yukawa-type quark operators, but the two processes are complementary in more complicated scenarios \cite{Roy:2025pht}. 
	If the invisible states are thermal dark matter, the achieved sensitivities are weaker than those from direct or indirect detection experiments, but they are complementary for more general dark matter scenarios and for certain parameter regions when many operators contribute at the same time.

	\section*{Acknowledgments}
	
	This work is supported by an Australian Research Council Discovery Project. SD is supported by the U.S. Department of Energy under Contract No. DE- SC0012704. 
	

	\section*{References}
	\bibliography{refs.bib}

\end{document}